\documentclass{iopart}
\usepackage{epsfig}

\def\Journal#1#2#3#4#5{#1 {\it #2} #3 {\bf #4} #5}
\def\Preprint#1#2{#1 {\it Preprint} #2}
\def\CPC{Comput. Phys. Commun.}
\def\JP{J. Phys.}
\def\NP{Nucl. Phys.}

\def\PR{Phys. Rev.}
\def\PRL{Phys. Rev. Lett.}
\def\PRT{Phys. Rep.}
\newcommand{\ro}{R_{\rm out}}
\newcommand{\rs}{R_{\rm side}}
\newcommand{\xo}{x_{\rm out}}
\newcommand{\xs}{x_{\rm side}}
\newcommand{\ptt}{p_{\rm T}}
\newcommand{\mt}{m_{\rm T}}
\newcommand{\sigp}{\sigma_p}

\begin{document}

\title{Kaon Interferometry at RHIC from AMPT Model}

\author{Zi-wei Lin \dag\footnote[3]{email: zlin@mps.ohio-state.edu} 
and C.M. Ko \ddag\footnote[5]{email: ko@comp.tamu.edu}}
\address{\dag\ 
Physics Department, The Ohio State University, Columbus OH 43210}
\address{\ddag\
Cyclotron Institute and Physics Department, Texas A\& M University, 
College Station, TX 77843-3366, USA}

\begin{abstract}
The two-kaon interferometry at RHIC is studied in a multi-phase transport
model. Similar to the pion case, we find strong space-time correlation at 
freeze-out for the kaon emission source, which results in a large positive 
$\ro\!-\!t$ term and tends to reduce the $\ro/\rs$ ratio. Unlike the pion 
case, the source radii for kaons determined from the emission function 
are close to the radius parameters extracted from a Gaussian fit 
to the correlation function. \footnote[9]{Slides of this conference talk at
{\it http://nt3.phys.columbia.edu/people/zlin/PUBLICATIONS/}}.  
\end{abstract}

\section{Introduction and Summary}

Particle interferometry based on the Hanbury-Brown Twiss (HBT) effect 
has been used extensively in heavy ion collisions to extract the information 
on the emission source of particles 
\cite{Pratt:su,Bertsch:1988db,Pratt:zq,Rischke:1996em}. 
In particular, the long emission time as a result of the phase 
transition from the quark-gluon plasma to hadronic matter in 
relativistic heavy ion collisions may lead to an emission 
source which has a much larger radius in the direction of the total 
transverse momentum of detected two particles ($R_{\rm out}$) than 
that perpendicular to both this direction and the beam direction 
($R_{\rm side}$) \cite{Rischke:1996em,Soff:2000eh}. 
Since the quark-gluon plasma is expected to be formed in heavy
ion collisions at RHIC, it is surprising to find that the extracted 
ratio $R_{\rm out}/R_{\rm side}$ from a Gaussian fit to the measured
two-pion correlation function in Au+Au collisions at $\sqrt s=130A$ GeV 
is close to one \cite{Adler:2001zd,Johnson:2001zi,Adcox:2002uc}, 
very different from predictions of hydrodynamical models 
\cite{Rischke:1996em,Soff:2000eh}. 

Since particle interferometry probes the phase-space distributions of
particles at freeze-out, it is natural to apply transport models to HBT. 
The reason is that particle freeze-out is dynamically generated
in transport models when the mean-free-path exceeds the system size 
at later stage of expansion, whereas freeze-out has to be imposed in 
hydrodynamical models. Using a multi-phase transport (AMPT) model, 
we have found that the small pion $\ro/\rs$ ratio could be due to a large
and positive space-time correlation in the emission source \cite{Lin:2002gc}. 
Furthermore, the pion source at freeze-out is highly non-Gaussian, leading 
to much larger pion source radii than the radius parameters from a Gaussian 
fit to the three-dimensional correlation function. 

In this study, we extend the work of Ref.\cite{Lin:2002gc} by studying 
the kaon interferometry in central Au+Au collisions at RHIC energies. 
Using the AMPT model, we find that, unlike the pion case, the kaon source 
radii extracted directly from the emission function are close to the fitted 
radius parameters extracted from a Gaussian fit to the three-dimensional 
correlation function. Our results also show that the kaon emission source 
has a large and positive correlation between time and position along 
the ${\rm out}$-direction at freeze-out, similar to what we have found 
earlier for the pion emission source. We expect that the study of 
kaon interferometry, as well as other observables such as the strange 
hadron elliptic flow, will be useful in understanding the dynamics of 
strange quarks and hadrons in heavy ion collisions at RHIC. 

\section{A MultiPhase Transport Model - AMPT}

The AMPT model is a hybrid model that uses the minijet partons 
from hard processes and excited strings from soft processes in the 
HIJING model \cite{Wang:1991ht} for the initial condition 
of relativistic heavy ion collisions.  The time evolution of 
partons is then described by the ZPC \cite{Zhang:1997ej} parton 
cascade model, and that of hadrons by an extended ART model \cite{Li:1995pr}. 
In the default AMPT model, only minijet partons are included in the parton 
cascade with a parton scattering cross section of $\sigma_p=3$ mb. After 
partons freeze out, they combine with their parent strings and then fragment 
to hadrons according to the Lund string fragmentation as implemented in 
PYTHIA \cite{Sjostrand:1993yb}. The default model has been quite reasonable 
in describing the measured rapidity distributions of charge particles 
\cite{Zhang:2000bd,Lin:2001cx}, particle to antiparticle ratios 
\cite{Lin:2001cx}, and the spectra of low transverse momentum pions, 
kaons \cite{Lin:2001yd}, multi-strange baryons \cite{Pal:2001zw} and $\phi$ 
mesons \cite{Pal:2002aw} in heavy ion collisions at SPS and/or RHIC. 

Since the initial energy density in Au+Au collisions at RHIC is expected 
to be much larger than the critical energy density at which the transition 
from hadronic matter to quark-gluon plasma would occur 
\cite{Zhang:2000nc,Karsch:2001vs}, the AMPT model has been extended to 
allow the conversion of initial excited strings to partons  
at RHIC energies \cite{Lin:2001zk,Lin:2002gc}. In this {\it string melting} 
scenario, hadrons that would have been produced from string 
fragmentation are converted instead to valence quarks and/or 
antiquarks. Interactions among these partons are again described 
by the ZPC parton cascade model. The transition from the partonic 
matter to hadronic matter at parton freeze-out is achieved 
using a simple quark coalescence model by combining two nearest 
partons into mesons and three nearest partons into baryons (or anti-baryons)
\cite{Lin:2001zk}. With the energy in excited strings taking part in 
the early partonic interactions and using quark coalescence to model
hadronization, the extended AMPT model with string melting \cite{Lin:2001zk} 
is able to describe the observed elliptic flow at RHIC 
\cite{Ackermann:2000tr,Adler:2002pu}, which the default AMPT model failed 
to reproduce. 

At present, the ZPC parton cascade \cite{Zhang:1997ej} in the AMPT model 
includes only two-parton elastic scatterings. The in-medium differential 
cross section is given by
$d\sigma_p/d\hat t=9\pi\alpha_s^2 (1+\mu^2/{\hat s})/2/(\hat t-\mu^2)^2$,
where the effective screening mass $\mu$ in principle depends on 
the temperature and density of the partonic matter\cite{Zhang:1997ej}. 
In this study, we take $\mu$ as a parameter in order to study 
the effect of partonic scatterings. Also, for simplicity, we assume the 
same scattering cross section for partons of different flavors. 
We note, however, that comparisons of high-quality data on the elliptic flow 
of strange hadrons \cite{Adler2,Sorensen:2003wi} 
with theoretical predictions 
\cite{Lin:2002rw,Voloshin:2002wa,Molnar:2003ff,Lin:2003jy,Greco:2003mm}
is expected to provide very useful information on the interactions of 
strange quarks in dense matter. 

\section{Two Ways of Extracting Radius Parameters}\label{khbt}

To evaluate the two-kaon correlation function requires the knowledge
of the single kaon emission function $S(x,{\bf p})$. In the AMPT model, 
it is obtained from the kaon space-time coordinate $x$ and momentum 
${\bf p}$ at kinetic freeze-out. The HBT correlation function 
for two identical hadrons of momenta ${\bf p_1}$ and ${\bf p_2}$ is then 
given by \cite{Pratt:su,Wiedemann:1999qn}
\begin{eqnarray}
C_2({\bf Q},{\bf K})=1+
\frac{\int   d^4x_1d^4x_2 S(x_1,{\bf K})
S(x_2,{\bf K}) \cos [Q\cdot(x_1-x_2)]}
{\int d^4x_1 S(x_1,{\bf p_1}) \int d^4x_2 S(x_2,{\bf p_2})},
\label{emission}
\end{eqnarray}
where ${\bf K}=({\bf p_1}+{\bf p_2})/2$ and $Q=({\bf p_1}-{\bf p_2}, E_1
-E_2)$. Expecting that the emission function is sufficiently smooth 
in momentum space, one can evaluate the correlation function by using
${\bf p_1}$ and ${\bf p_2}$ for ${\bf K}$ in the numerator of above
equation.

First, the size of the emission source can be extracted from the 
emission function via the curvature of the correlation function 
at ${\bf Q}=0$:
\begin{equation}
R_{ij}^2(K)
=D_{x_i,x_j}(K)-D_{x_i,\beta_j t}(K)-D_{\beta_i t,x_j}(K)+
D_{\beta_i t,\beta_j t}(K). 
\label{source}
\end{equation}
In the above, ${\bf \beta}={\bf K}/K_0$ with $K_0$ denotes the average 
energy of the two kaons; $x_i(i=1-3)$ are spatial coordinates of a kaon 
at freeze-out; and $D_{x,y}=\langle x\cdot y \rangle-\langle x\rangle 
\langle y \rangle$ with $\langle x\rangle$ denoting the average value 
of $x$. In this study we use the usual ``out-side-long'' (${\it osl}$) 
coordinate system \cite{Bertsch:1988db,Pratt:zq}. 

On the other hand, the measured correlation function $C_2({\bf Q},{\bf K})$ 
is usually fitted by a Gaussian function in ${\bf Q}$, i.e., 
\begin{eqnarray}
C_2({\bf Q},{\bf K})=1+ \lambda \exp 
\left ( -\sum_{i=1}^3 R^2_{ii}(K) Q_i^2 \right ),
\label{gaussian}
\end{eqnarray}
We note that, for central heavy-ion collisions, the above fitted radius 
parameters would be identical to the source radii given by the curvature of
the emission function at ${\bf Q}=0$ (i.e. Eq.~(\ref{source}))  
only when the emission source is Gaussian in space and time. 

\section{AMPT Results at RHIC Energies}

\subsection{Two-kaon correlation function}

\begin{figure}[ht]
\centerline{\epsfig{file=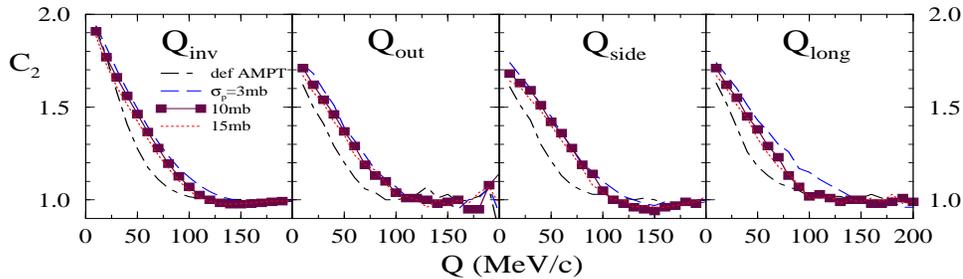,width=5in,height=1.5in,angle=0}}
\caption{Correlation functions for $K^0_S$ ($-1<y<1$, $200<\ptt<400$ MeV$/c$).}
\end{figure}

Using the program Correlation After Burner \cite{pratt:uf}, we have 
evaluated the correlation function $C_2({\bf Q},{\bf K})$ of two 
kaons in their longitudinally comoving frame for central ($b=0$ fm) Au+Au 
collisions at $\sqrt{s}=130A$ GeV. In Figure~1, we show the 
invariant correlation function, i.e., the correlation function 
as a function of $Q_{\rm inv} \equiv \sqrt{-Q^2}$, and its projections 
onto one of the $Q_{\rm out}, Q_{\rm side}$ and $Q_{\rm long}$ axes. 
In evaluating the projected correlation function, the other two 
${\bf Q}$-components have been integrated over the range $0-40$ MeV/$c$. 
The dash-dotted curves in Figure~1 represent results from the default AMPT 
model (no string melting), while the other curves are from the extended AMPT 
model with string melting but using different values for $\sigma_p$. 
It is seen that the one-dimensional kaon correlation functions become 
narrower as $\sigp$ is increased in the extended AMPT model. Their 
dependence on $\sigp$ seems, however, to be much weaker than that of 
the pion correlation function \cite{Lin:2002gc}. 

\subsection{Source radii versus fitted radius parameters}

The source radii for kaons within $-1<y<1$ and $200<\ptt<400$ MeV$/c$
are shown in Figure~2 (a). The results are obtained from both the default 
AMPT model (shown at $\sigp=0$) and the extended AMPT model with string 
melting. For the latter, we have used different parton cross sections of 
$\sigp=3$, 6, 10, and 15 mb. It is seen that these radii have values 
between 2 and 5 fm, and they are much smaller than the source radii for 
low $\ptt$ pions (between 7 and 25 fm) \cite{Lin:2002gc}. The $\ro/\rs$ 
ratio (solid curves without symbols) from the kaon emission function is 
also shown in Figure~2 (a), and it increases appreciably with increasing 
parton scattering cross section.

\begin{figure}[ht]
\begin{center}
\begin{tabular}{lr} 
\epsfig{width=0.46 \columnwidth, height=0.36 \columnwidth, 
file=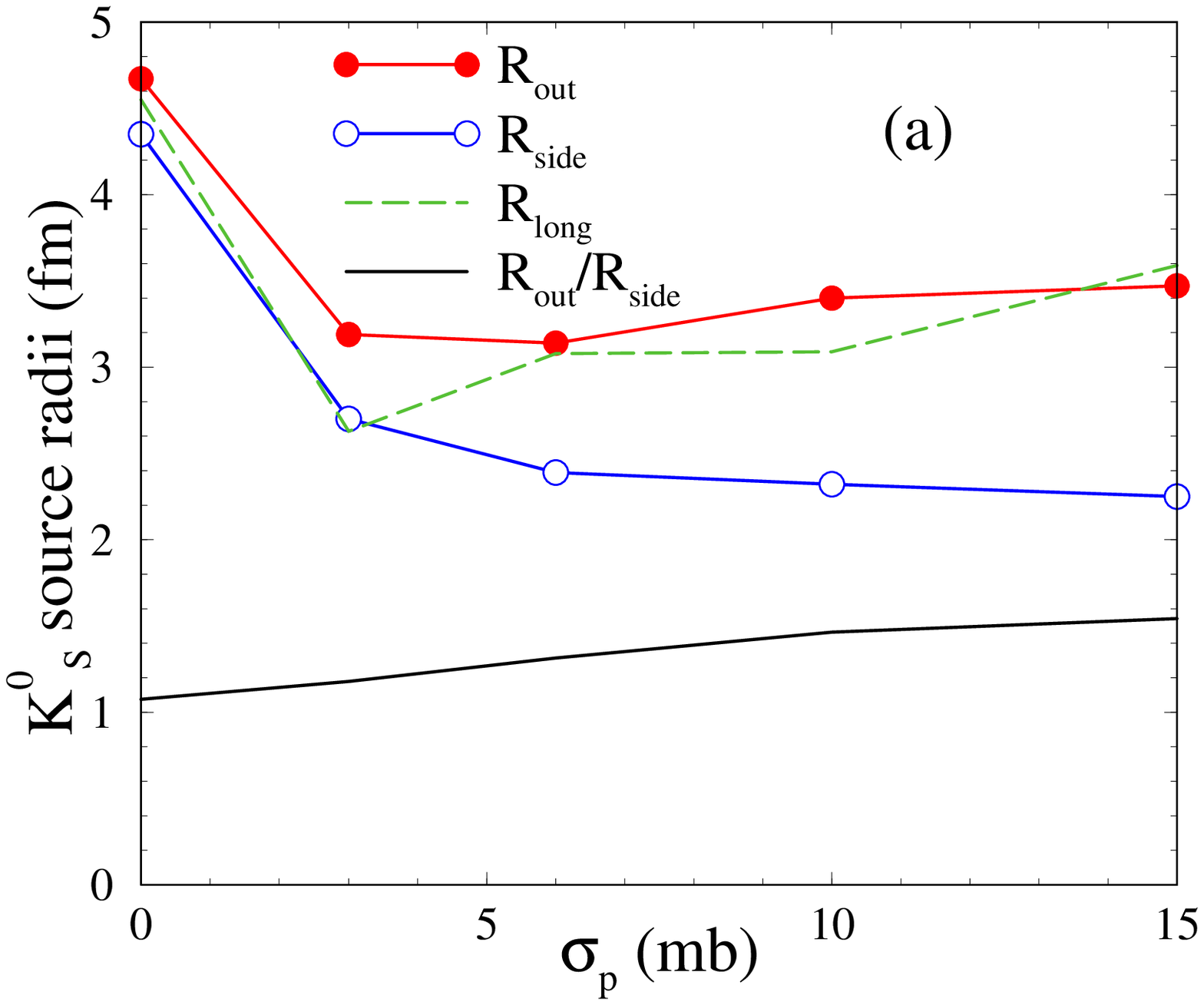}
&
\epsfig{width=0.46 \columnwidth, height=0.372 \columnwidth, 
file=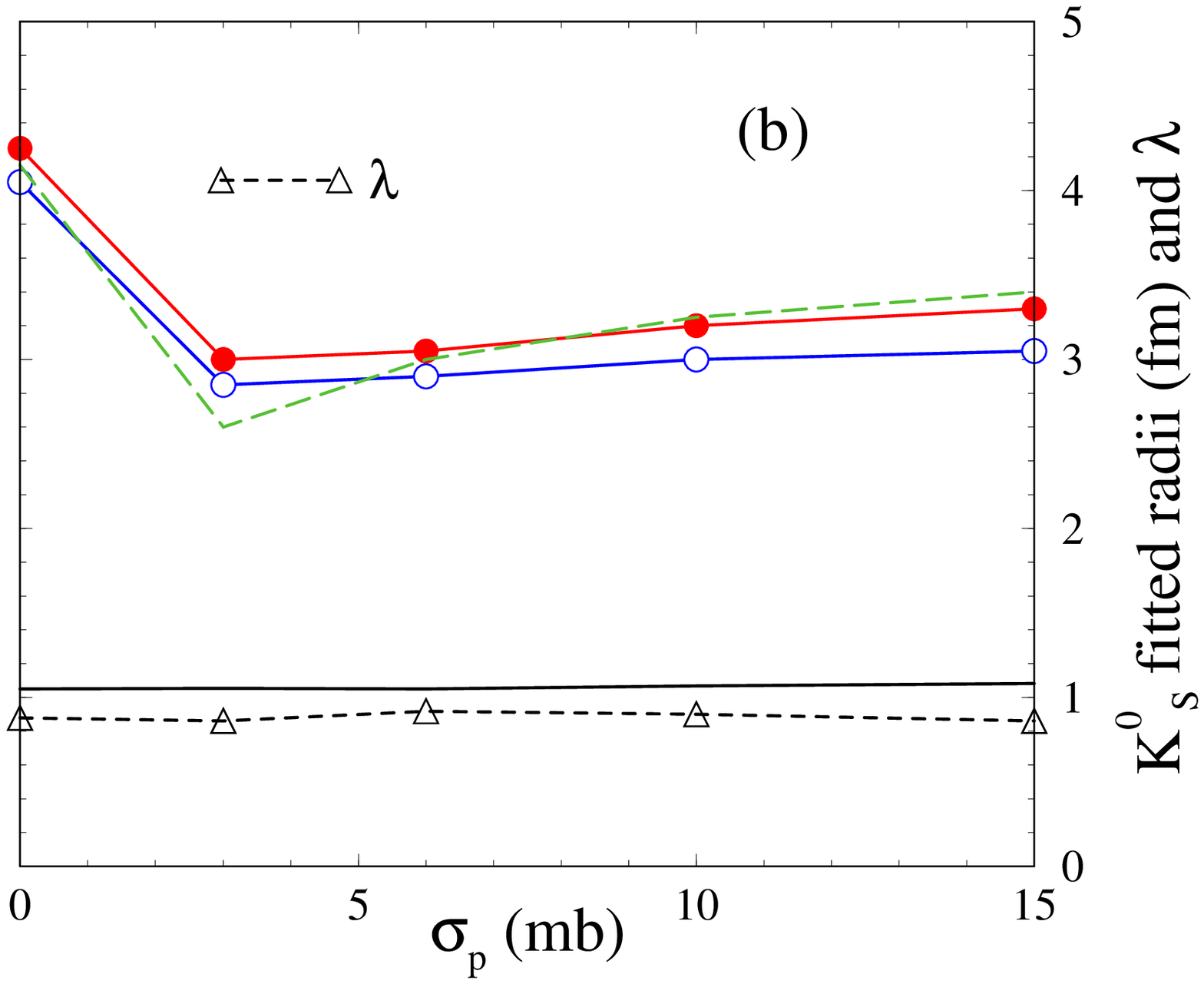}
\end{tabular}
\caption{(a) Source radii, (b) fitted radius parameters and $\lambda$ 
for $K^0_S$ as functions of $\sigp$ at $130A$ GeV. Points at $\sigp=0$ 
correspond to the default AMPT.
}
\end{center}
\end{figure}

The radius parameters obtained from fitting the three-dimensional 
correlation function $C_2({\bf Q})$ by Eq.~(\ref{gaussian}) are shown
in Figure~2 (b). We find that they are close (mostly within $30\%$)
to the source radii determined directly from the emission function.
This is contrary to the pion case, where the pion source radii 
can be more than twice larger than the radius parameters from the 
Gaussian fit \cite{Lin:2002gc}. Note that kaons from $\phi$ meson decays 
have been included in the evaluation of the correlation function 
$C_2({\bf Q})$ but not in the source radii (due to the long lifetime of 
$\phi$); however, pions from $\omega$ decays have been included in the 
source radii in Ref.~\cite{Lin:2002gc}. Figure~2 (b) also shows that 
$\ro/\rs$ from fitted radius parameters changes little with parton 
cross section, in contrast to the significant increase seen in the source 
radii shown in Figure~2 (a).  

\begin{figure}[ht]
\begin{center}
\begin{tabular}{lr} 
\epsfig{width=0.44 \columnwidth, height=0.36 \columnwidth, 
file=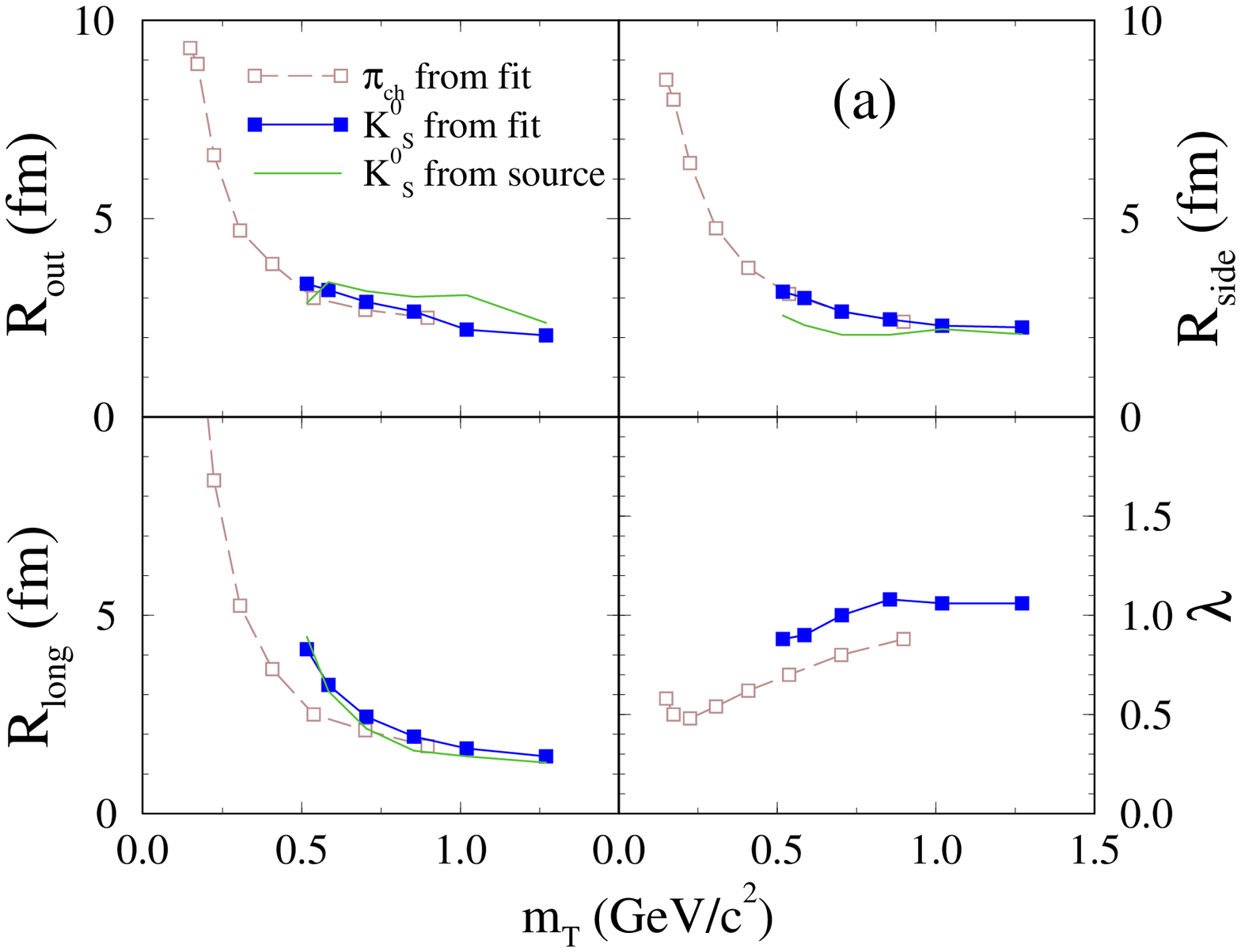}
&
\epsfig{width=0.5 \columnwidth, height=0.3 \columnwidth, 
file=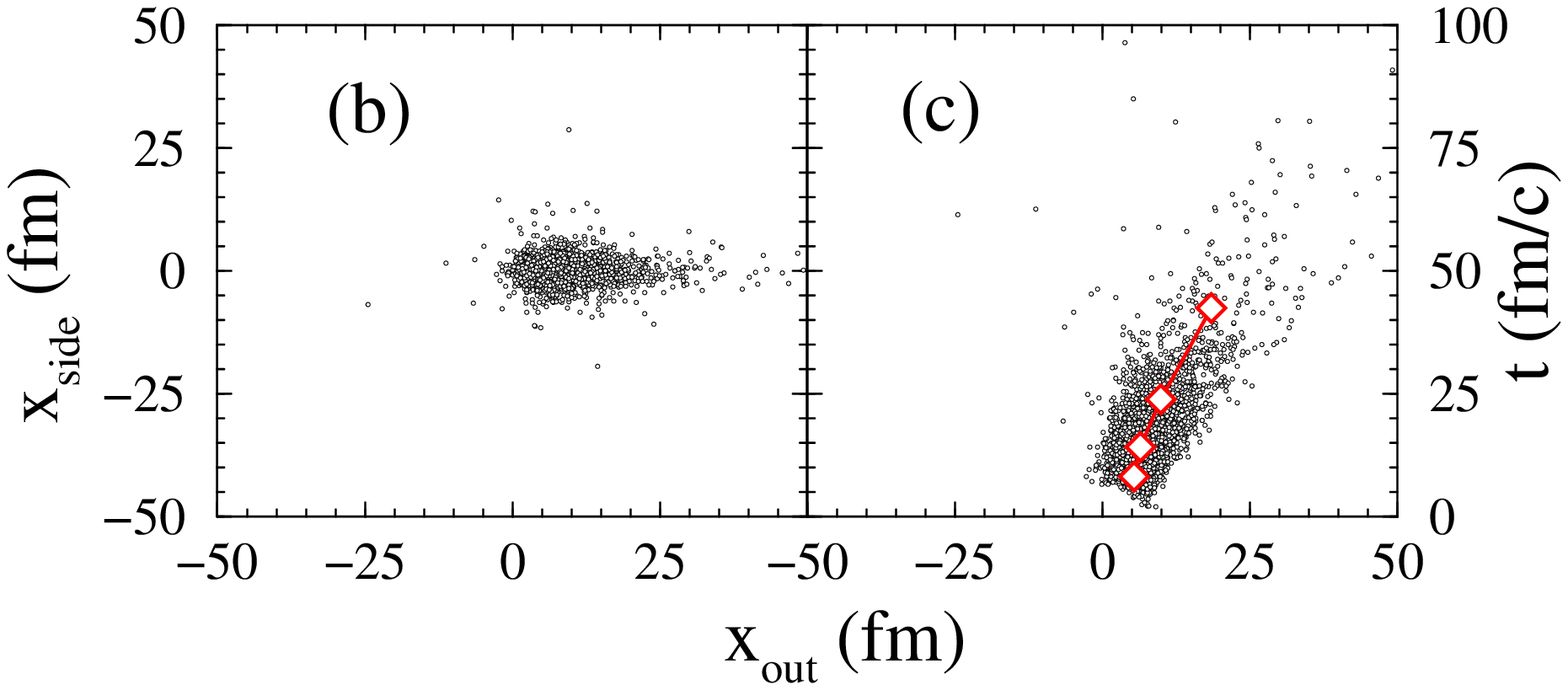}
\end{tabular}
\caption{(a) Source radii and fitted radius and $\lambda$ parameters 
for $K^0_S$ with $-1<y<1$ as functions of $m_{\rm T}$, as well as 
fitted radius and $\lambda$ parameters for mid-rapidity pions 
\cite{Lin:2002gc}. (b) $x_{\rm out}-x_{\rm side}$ and (c) $x_{\rm out}-t$ 
distributions at freeze-out for $K^0_S$. The curve with open diamonds 
represents $\langle x_{\rm out} \rangle(t)$.
}
\vspace{-0.1cm}
\end{center}
\end{figure}

Figure~3 (a) shows the extended AMPT results with $\sigp=10$ mb for
kaon source radii as functions of $\mt$ in the six $\ptt$-bins, 
$0-200-400-600-800-1000-1500$ GeV$/c$, for Au+Au collisions 
at $\sqrt s=130A$ GeV. We see that for all $\mt$ the source radii (solid) 
are close to the corresponding fitted radius parameters (filled squares).
We note that, before making quantitative predictions on kaon correlations, 
abundances of strange resonances such as $K^*$ and $\phi$ need to be checked, 
and the effects of other resonances not yet included in the AMPT model 
also need to be investigated. 

\subsection{$x-t$ correlation and the $\ro/\rs$ ratio}

Figure~3 (b) and (c) show the $\xo-\xs$ and $\xo-t$ distributions 
at freeze-out for $K^0_S$ with $200<\ptt<400$ MeV$/c$ and $-1<y<1$. 
The kaon emission source shows a positive shift in $\xo$ as in the pion case 
\cite{Lin:2002gc}, consistent with a strong transverse flow. The emission 
source also appears to be highly non-Gaussian, which leads to different 
values of radius parameters extracted from the two methods in 
Section~\ref{khbt} \cite{Hardtke:1999vf,Lin:2002gc}.
The solid curve with open diamonds in Figure~3 (c) gives the average value 
$\langle x_{\rm out} \rangle$ as a function of freeze-out time $t$. 
The kaon emission source is seen to have a strong positive $x_{\rm out}-t$ 
correlation as in the pion emission source \cite{Lin:2002gc}.

Since $\rs^2={D_{\xs,\xs}}$, but 
\begin{equation}
R_{\rm out}^2=D_{x_{\rm out},x_{\rm out}}-2~D_{x_{\rm out},\beta_\perp t}
+D_{\beta_\perp t,\beta_\perp t}, 
\label{rout}
\end{equation}
the ratio $R_{\rm out}/R_{\rm side}$ contains information about 
the duration of emission (in the last term) and has been studied extensively 
\cite{Rischke:1996em,Bernard:1997bq,Soff:2000eh,Soff2}. 
We note, however, that a direct relation between $R_{\rm out}$ and the 
emission duration exists only if the $x_{\rm out}-t$ correlation term 
$D_{x_{\rm out},\beta_\perp t}$ is small, which is not the case according 
to our results from the extended AMPT. E.g, the above equation for the 
kaons is numerically written as $3.4^2 \simeq 35-2\times22+20$ in units 
of fm$^2$. The $x_{\rm out}-t$ correlation term 
$D_{x_{\rm out},\beta_\perp t}$ is thus positive ($+22$ fm$^2$) and 
comparable to the magnitude of $D_{\beta_\perp t,\beta_\perp t}$ 
($20$ fm$^2$), making it difficult to extract information about the duration 
of emission from $R_{\rm out}/R_{\rm side}$. The situation is similar 
for the pion emission source, e.g., for mid-rapidity pions with 
$125<\ptt<225$ MeV$/c$, numerically the corresponding Eq.~(\ref{rout}) is 
$17^2 \simeq 185-2\times 168+431$ in units of fm$^2$. 

\subsection{Energy dependence}

Figure~4 shows the correlation functions for both neutral and charged kaons 
within $-1<y<1$ and $200<\ptt<400$ MeV$/c$ from AMPT model with string 
melting and $\sigp=10$ mb for $\sqrt s=130A$ and $200A$ GeV. 
The effect due to Coulomb interactions is included for $K^+$ 
correlation functions using the program Correlation After Burner 
\cite{pratt:uf}. The kaon correlation functions are found to change only 
slightly from $130A$ to $200A$ GeV at RHIC.

\begin{figure}[ht]
\centerline{\epsfig{file=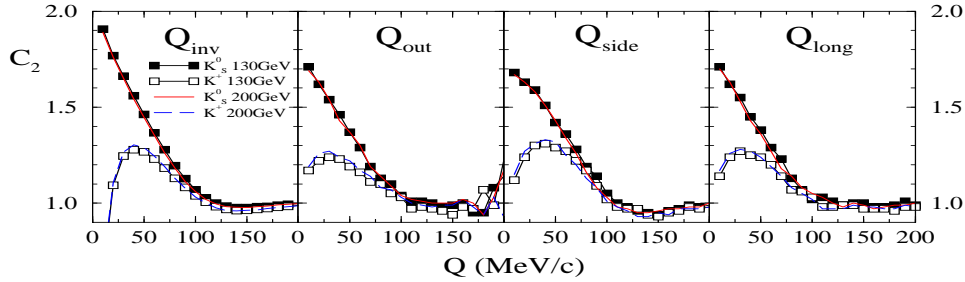,width=5in,height=1.5in,angle=0}}
\caption{Correlation functions for $K^0_S$ and $K^+$ at $130A$ and $200A$ GeV. 
}
\vspace{-0.1cm}
\end{figure}

\section*{Acknowledgments} 
We appreciate stimulating discussions with H. Huang, M. Murray, S. Pratt 
and N. Xu. This work was supported by the U.S. Department of Energy Grant 
No. DE-FG02-01ER41190 (Z.W.L.) and by the U.S. National Science Foundation
Grant No. PHY-0098805 as well as the Welch Foundation under Grant No. A-1358 
(C.M.K.)

{}
\end{document}